\documentclass[]{spie}  

\usepackage{amsmath,amsfonts,amssymb}
\usepackage{graphicx}
\usepackage[colorlinks=true, allcolors=blue]{hyperref}
\usepackage{soul}

\title{Optical development of the BISOU breadboard}

\author[a,b]{*Morgane Loquet Le Gall}
\author[c]{Creidhe O'Sullivan}
\author[a]{Bruno Borgo}
\author[a]{Clémence De Jabrun}
\author[a]{Valentin Sauvage}
\author[c]{Neil Trappe}
\author[a]{Bruno Maffei}
\affil[a]{Institut d'Astrophysique Spatiale, Université Paris-Saclay, CNRS, Building 121, Orsay, France}
\affil[b]{Centre national d'études spatiales (CNES), 75039 Paris, France}
\affil[c]{Department of Physics, Maynooth University, Co. Kildare, Ireland}

\authorinfo{Further author information: (Send correspondence to M. Loquet Le Gall) \\ E-mail: morgane.loquet-le-gall@universite-paris-saclay.fr}

\begin{document} 
\maketitle

\begin{abstract}
BISOU (Balloon Interferometer for Spectral Observations of the primordial Universe) is an astronomical balloon-borne pathfinder developed as part of a preparatory study for a future space mission aimed at measuring spectral distortions of the cosmic microwave background (CMB). A laboratory breadboard of the instrument is being developed at the Institut d’Astrophysique Spatiale (IAS), enabling the characterization of subsystems and instrument systematic effects, particularly in the optical system. The optical system is based on a differential polarizing Fourier Transform Spectrometer (FTS) that receives inputs from both a sky-facing telescope and an internal calibration source. The FTS focal planes include sub-K detectors coupled to multimode feed horns. The full spectral band, spanning between 90 and 1500 GHz, is sub-divided into two frequency sub-bands, thanks to the use of a dichroic. The optical analysis first relies on ray-tracing simulations to establish the overall configuration of the system, before proceeding to more advanced Gaussian beam and physical optics analyses.
\end{abstract}



\section{INTRODUCTION} 

The last and almost unique measurement of the cosmic microwave background (CMB) spectrum was performed by the COBE-FIRAS\cite{1990ApJ...354L..37M} space mission. Although the CMB is well described by a perfect blackbody at a
temperature of 2.725 ± 0.00057 K \cite{CMBtempfixsen}, theoretical model predicts small deviations, referred to as CMB spectral distortions \cite{spectral_distortions}. They could have been produced by a wide range of physical processes leading to energy injection in the early and late-time Universe; the thermal Sunyaev-Zel’dovich effect being one example. Depending on the time at which energy is injected into the primordial plasma, distortions take characteristic spectral shapes. Measuring these distortions is an important probe of the thermal history of the Universe but, due to the limited sensitivity of FIRAS, only upper limits have been established. 

\noindent FOSSIL (FTS fOr CMB Spectral diStortIon expLoration) \cite{FOSSIL} is a concept space mission proposed in response to the ESA M8 call, targeting the measurement of $\mu$- and $y$-distortions at a sensitivity level far beyond FIRAS limits. Such a level of precision can only be reached if the instrument's systematics are well known and controlled. Studying the systematic effects of the measurement concept is one of the aims of BISOU \cite{BISOU} (Balloon Interferometer for Spectral Observations of the primordial Universe), its main scientific objectives being the first measurement of the $y$-monopole distortion and a more accurate measurement of the Cosmic Infrared Background (CIB) \cite{CIB}.


\noindent While FOSSIL remains a concept mission and BISOU has yet to be selected and built (presently in a CNES Phase A study), the development of both a cold and a warm breadboard model (BBM) represents a critical step towards the consolidation of their measurement concept and the characterisation of its unknown systematics. In particular, the asymmetry of the two FTS optical paths and the cryostat's window emissivity will need to be carefully studied. Beyond validating the overall optical design, the breadboard will also serve as a test of new technologies, including detectors and the dichroic component.

\noindent This paper presents the measurement concept of BISOU and its associated breadboard model, the cryogenic test facility developed to validate the design, and the current optical design of the instrument.

\section{The BISOU balloon mission}

\subsection{Measurement principle}
\label{measurement principle}

\begin{figure}[h!]
    \centering
    \includegraphics[width=0.75\linewidth]{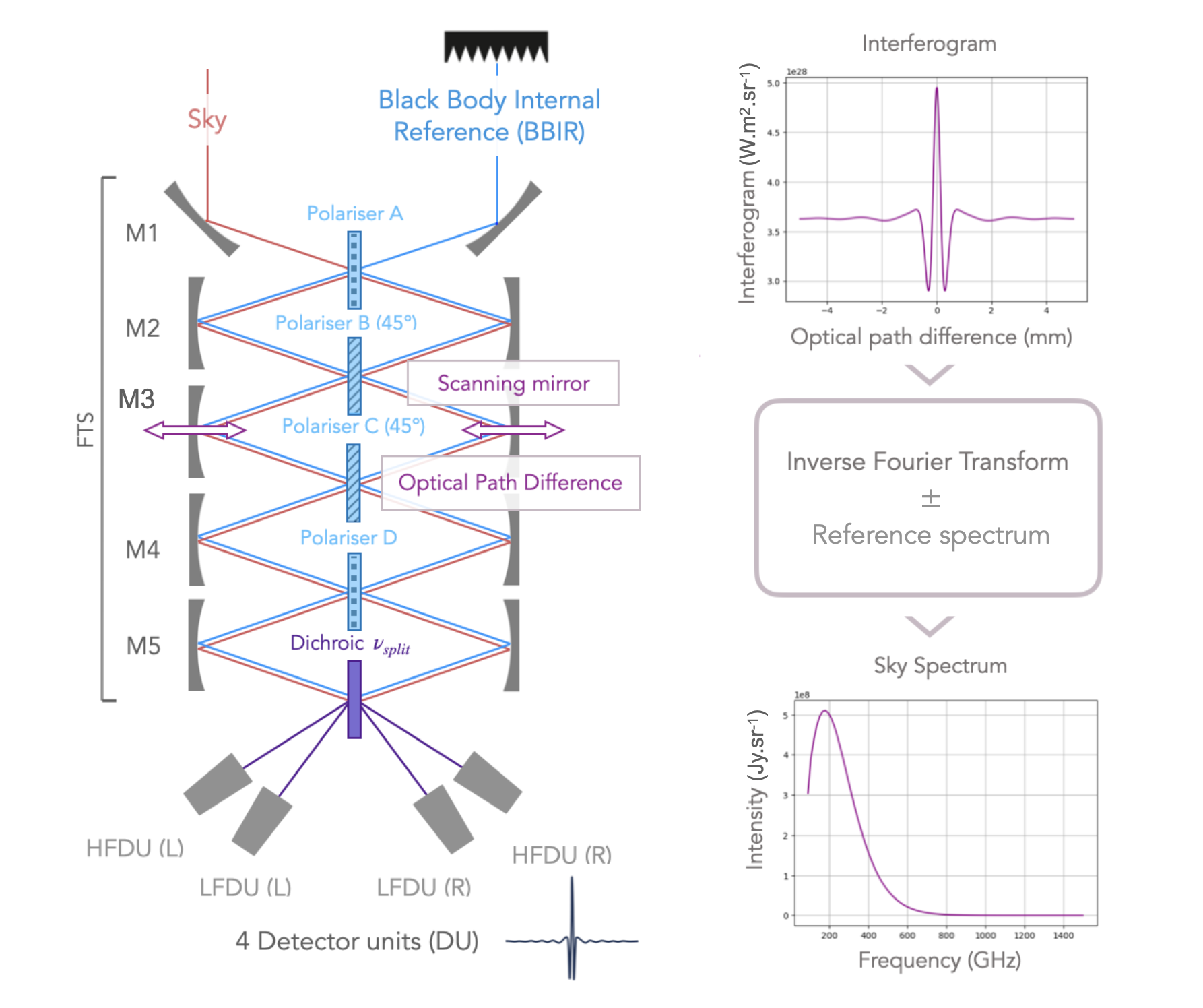}
    \caption{\textbf{Left}: Conceptual view of the measurement method. The sky signal (red) and the Black Body internal reference (BBIR, blue) are injected into the interferometer by mirrors M1, then combined through a sequence of polarisers (A,B,C and D). Mirrors M3 introduce an optical path difference (OPD) between the two arms of the FTS. At the output, a dichroic splits the signal into two spectral bands (high and low frequency), measured by four detector units: HFDU and LFDU, for the left (L) and right (R) channels.
    \textbf{Right}: Illustration of the spectrum reconstruction. The  interferogram, measured as a function of OPD is processed via Fourier transform and subtracted/added to the reference spectrum, to reconstruct the sky spectrum.}

    \label{fig:measurement_principle}
\end{figure}

The measurement of the CMB spectral distortions requires an absolute comparison between the sky and a well-known reference. For this purpose, the instrument concept (see Fig. \ref{fig:measurement_principle}) is based on a polarised Martin-Puplett Fourier Transform Spectrometer \cite{MPInterferometer} (FTS) with two inputs. One arm of the FTS views the sky while the other looks at an internal blackbody reference maintained at the temperature of the CMB (2.7~K). The beams are split and recombined using wire-grid polarisers, and a scanning mirror introduces an optical path difference (OPD) $\delta$ between the two inputs to produce an interferogram. The outputs of the FTS are coupled to detectors through multimode feedhorns. The measured signal consists of a constant part and a scan-modulated part that is proportional to the difference between the two inputs. Applying an Inverse Fourier Transform and adding or subtracting the known spectrum of the reference allows for sky spectrum retrieval. Finally, a dichroic is added before the focal plane to split the wide frequency range into two sub-bands, improving the sensitivity of the instrument and resulting in low frequency detection units (LFDU) and high frequency detection units (HFDU).

\subsection{Application to the BISOU instrument}

With a frequency range from 90~GHz to 1.5~THz and a spectral resolution of 15~GHz, the BISOU instrument design follows the measurement principle presented in section \ref{measurement principle} but is adapted to meet the constraints of a balloon-borne experiment. It faces additional challenges that are not present for a traditional space mission. 

\begin{figure}
    \centering
    \includegraphics[width=0.5\linewidth]{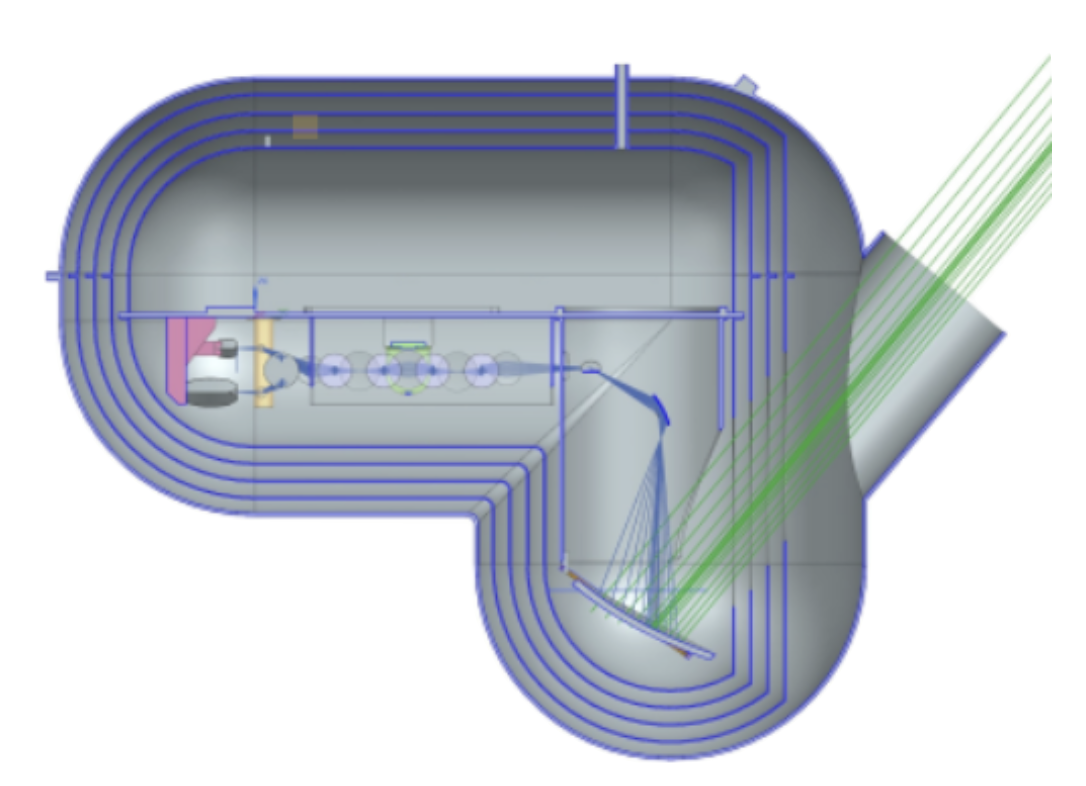}
    \caption{Cross-sectional view of the BISOU CAD model. The helium tank is visible above the cold plate, on
which the FTS is mounted.}
    \label{fig:CAD}
\end{figure}

The presence of residual atmosphere at its float altitude (40~km) requires the instrument to be enclosed in a vacuum-tight cryostat in order to maintain it at a very low temperature T$\sim$2.7~K (Fig.\ref{fig:CAD}). The optical entrance aperture will be closed by a window, transparent to millimeter and sub-millimeter waves, which, at this altitude, will be at a temperature of $\sim$ 270~K. Therefore, the thermal emission of this warm optical component will dominate the photon noise. Thermal filters located on each thermal shield will limit the thermal load on the cryogenic stages of the instrument and on the detectors.

\begin{figure}
    \centering
    \includegraphics[width=0.9\linewidth]{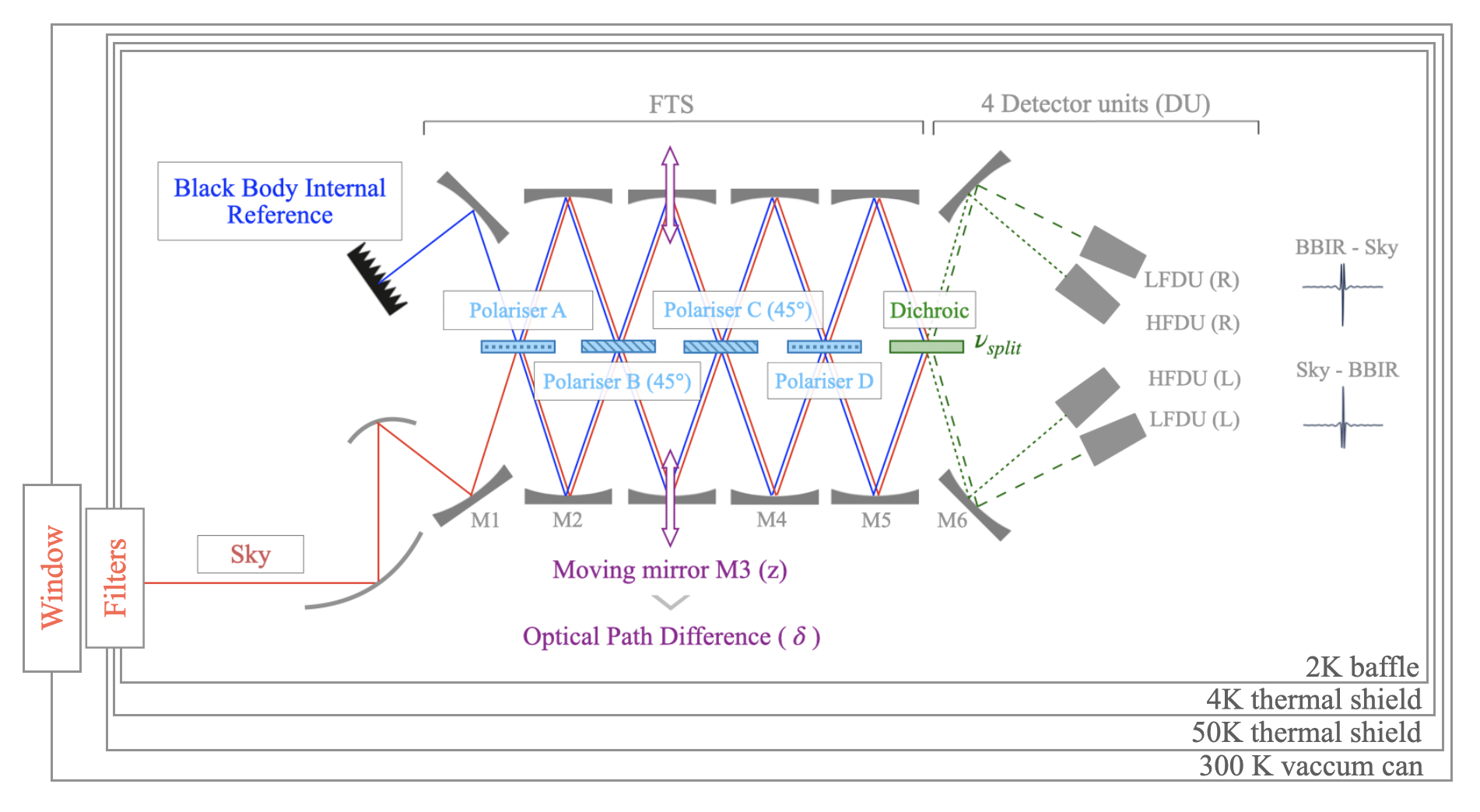}
    \caption{Schematic diagram of the BISOU Breadboard . One beam comes from the sky and another from the internal 2.7 K reference. An optical path difference is introduced by the pair of moving mirrors. Polarisation is split using polarising filters, and then recombined to produce the desired interference pattern. Before the focal plane, a dichroic separates the high and low frequencies, producing four focal planes. Each detector will be coupled to a multimode feedhorn optimised for its respective frequency band. The instrument is placed inside the cryostat environment and is looking at the outside through windows and filters.}
    \label{cryostat_schematic}
\end{figure}

\noindent While one arm of the FTS is looking at the sky through a telescope, the other arm is facing the internal reference directly (Fig.\ref{cryostat_schematic}). This asymmetry in the optical path needs to be studied since the additional elements and reflections introduced by the telescope can add cross-polarisation effects and aberrations that will not be present in the internal reference arm.

\noindent To improve the sensitivity of the instrument, the wide frequency band is split into two sub-bands using a dichroic filter with a cut-off frequency of about 300~GHz. Frequencies below this threshold are transmitted through the dichroic, while higher frequencies are reflected. This flux division significantly reduces the optical power incident on the low frequency band detectors, lowering the photon noise \cite{spiexavier}. This is an important consideration for a balloon-borne FTS instrument, where high-frequency emissions are dominated by the thermal emission of warm instrument components, together with the residual atmosphere. The resulting reduction in photon noise directly translates into an improved sensitivity in the low-frequency channel, where the specific spectral distortion signature needs to be measured. \cite{coulon2026spie}

\noindent All these optical specificities and potential sources of systematic effects must be carefully characterised, which is precisely the purpose of the breadboard model described in the following sections.

\section{BreadBoard Model Optical design}

\subsection{Experimental facility}

As described in the previous section, the breadboard model aims to study systematic effects and validate technologies prior to their integration into the BISOU and FOSSIL designs. The BBM is therefore designed to be highly reconfigurable, allowing each component of the instrument to be characterized.

The FTS will be placed on a 2~K cold plate inside a cryogenic test facility in order to test the components at the temperature at which they will operate in flight. The cryogenic chamber environment is composed of different shields at 300~K, 50~K and 4~K. In order to reach the operating temperature of the detectors, the focal plane will be cooled to 100~mK or 50~mK (depending on the chosen technology) by a dilution refrigerator from Bluefors (model SD250). Integrated within the breadboard cryostat facility, it provides 250~$\mu$W of cooling power at 100~mK and can reach temperatures as low as 30~mK, which is more than sufficient. The 2~K cold plate is cooled by the 30~mW of cooling power available on the still stage of the SD250. The 2~K baffle is thermally connected to the cold plate. The 4~K shield and the 50~K shield are cooled by a Cryomech PT415-RM (1.35~W at 4.2~K and 40~W at 45~K). The FTS motor will be located on a 4~K plate attached beneath the 2~K cold plate via insulating supports (G-10) and cooled by the 4~K stage. 

A key feature of the cryostat will be its 200~mm diameter window on the 300~K vacuum can. Filters will be placed on the 50~K and 4~K thermal shields, allowing us to study their systematic effects on the measurements. This aperture will allow us to couple the cryostat to an atmospheric chamber as an input in order to understand how the atmospheric variations affect the measurements. It will also allow us to use the cryogenic facility for the calibration of the final BISOU instrument. In addition, the controlled thermal environment will enable the study of systematic effects induced by thermal gradients within the optical components.

\subsection{Optical schematic}

The optical design of BISOU relies on three complementary approaches. Ray tracing, performed with Zemax software \cite{zemax}, provides fast geometrical modelling to select mirror configurations and satisfy all design constraints. Gaussian beam optics \cite{goldsmith} then enables analytical estimation of the beam size at each optical element, which is used to optimise mirror dimensions, their illumination, and define edge taper targets. Full physical optics simulations are carried out with GRASP \cite{grasp} by propagating Maxwell's equations through the design, capturing diffraction effects and beam deformation, which become critical at the lowest frequencies where the beam is widest.

\begin{figure}
    \centering
    \includegraphics[width=1\linewidth]{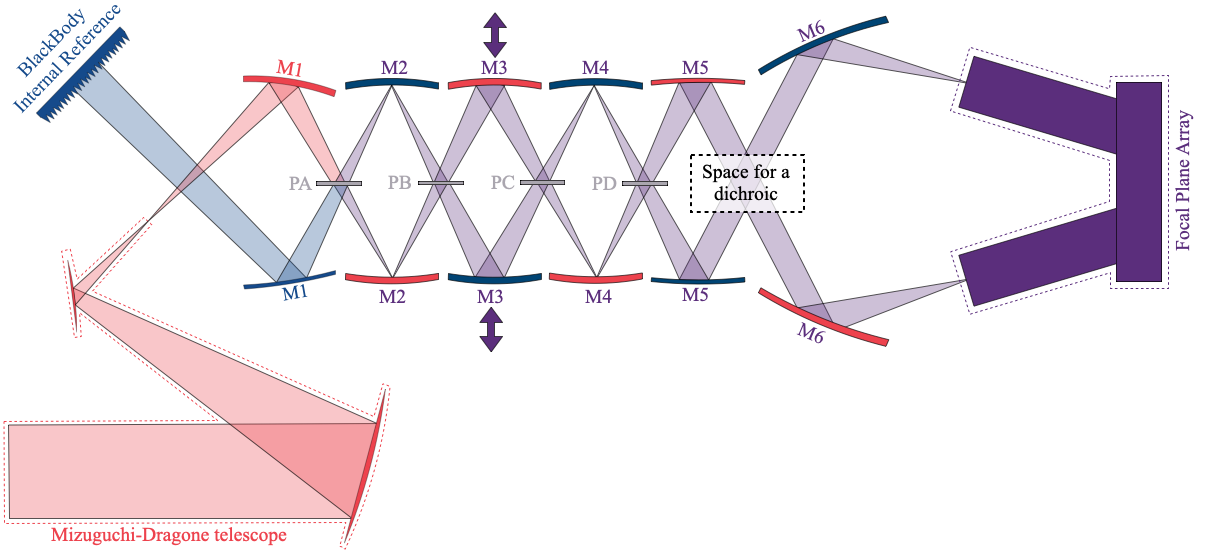}
    \caption{Optical layout of the BBM instrument. Light is collected from the BlackBody Internal Reference source (top left) or entered by the off-axis Mizuguchi-Dragone telescope (bottom left). The beam is then directed into a relay optical chain composed of five pairs of curved mirrors (M1–M5). The two optical paths (reference side and sky side) are represented in blue and red, respectively. Mirror M3 is mounted on a scanning mechanism allowing modulation along the optical axis. A space for the dichroic element is reserved between M5 and M6. The beam is finally focused by a pair of mirrors (M6) onto the Focal Plane Array, where the detectors are located.}
    \label{fig:BBm}
\end{figure}

The FTS consists of five pairs of mirrors, numbered from M1 to M5, guiding the beam through polarisers. On the telescope side colored in red in Fig~\ref{fig:BBm}, mirrors M1 to M4 are (M3 to M6), while on the reference side (represented in blue), only M2 to M4 are ellipsoidal. M1 remains a parabolic mirror as it is facing the internal calibrator reference. M3 mirrors are the moving element, translated by the motor. The dichroic is one of the most critical components of the design. It is placed in a collimated beam in order to avoid any spurious effect which might arise from incident rays reaching the dichroic surface with different incidence angles. For this purpose, the last FTS mirror M5 has been changed from an elliptical mirror, to a parabolic mirror. The detectors are required to operate at sub-K temperature, so in order to optimise the thermo-mechanical design the detection units (detectors and their associated feedhorns) must be co-located within the same area to form a single Focal Plane. This requirement is achieved through the use of the M6 mirrors, re-directing and focusing the beams onto the feedhorns.

Wire grid polarisers are located in between the path of the FTS mirrors. The wires reflect the parallel component of the incident beam while transmitting the perpendicular one.
The polarisers are used to combine the beams from the two sources: the sky and the internal reference. Polarisers A and D are oriented with their wires vertical (0°), while polarisers B and C are rotated by 45°.

The telescope follows the Mizuguchi-Dragone \cite{MD_condition} condition in its off-axis Cassegrain configuration, minimizing cross-polarization and astigmatism induced by reflections on the mirrors. The design was first established through ray tracing, then refined using Gaussian beam propagation to meet the -20~dB edge taper requirement on the 150~mm diameter primary reflector. 

The design presented in Fig.~\ref{fig:BBm} corresponds to the current warm model configuration. 
Compared to the BISOU instrument and the cold breadboard, both including four Detector Units (DUs), the warm model is reduced to two DUs. This simplified configuration allows us to first characterize the optics and perform the optical alignment more easily, as the feedhorns lie in the FTS plane. In the 4-DUs configuration, the feedhorns and their associated detectors are arranged in a two-level layout, placing them outside the FTS plane. The focal plane is designed to be modular, enabling different configurations to be tested.

\subsection{Future work}

Additional work is planned in the near future. Gaussian beam simulations of the full instrument will be carried out to further optimize the optical design. For instance, specific studies on the de-pointing, the aberrations and the spillover induced by the movement of the M3 moving mirror will be investigated, even though, the analysis carried out on the BISOU instrument \cite{spieopticaldesign} showed that moving M3 with a translation of four times the real mirror stroke does not introduce major impacts on the beam shape. The Gaussian beam simulation will then be improved by taking into account the multimode behaviour of the feedhorn to model the effective beam shape of the instrument. Moreover, special attention will be paid to the accurate calculation of the optical path difference (OPD), a critical parameter in the analysis and reconstruction of the data from the interferogram. Indeed, any systematic error in the OPD determination will introduce artificial distortions in the reconstructed spectrum.\\

The construction of the warm breadboard model is expected to be completed by the end of 2026, marking a key milestone toward the validation of the instrument concept.

\section{Conclusion}

In this work, we have presented the optical design of the BISOU breadboard model, developed at IAS as a critical step toward the realisation of the BISOU stratospheric balloon project and the future FOSSIL space mission. The measurement concept, based on a differential polarising Martin-Puplett FTS with two inputs, has been described alongside the cryogenic test facility that will enable the characterisation of the instrument at its operating temperature. The warm breadboard model, currently under construction, will allow for the validation the optical design and alignment procedures before the cold model is assembled.

\acknowledgments 
 
The authors acknowledge financial support from the Centre national d’études spatiales (CNES), France (ROR: https://ror.org/04h1h0y33), within the framework of the BISOU balloon project.

This research is supported by R\'egion \^Ile-de-France through fundings with reference IDF-DIM-ORIGINES-2023-4-07 and IDF-DIM-ORIGINES-2024-1-06
\bibliography{report} 
\bibliographystyle{spiebib} 

\end{document}